\documentclass[useAMS,usenatbib]{mn2e}

\usepackage{graphicx,amssymb}

\newcommand{\plotwd}{8.5cm}

\title[Power spectrum of the maxBCG cluster sample: new evidence for the acoustic features]{Power spectrum of the maxBCG cluster sample: new evidence for the acoustic features}
\author[G. H\"utsi]{G. H\"utsi$^{1,2}$\thanks{E-mail: ghutsi@star.ucl.ac.uk}\\
$^{1}$Department of Physics and Astronomy, University College London, London, WC1E 6BT\\
$^{2}$Tartu Observatory, EE-61602 T\~oravere, Estonia}
\begin{document}

\date{}

\pagerange{\pageref{firstpage}--\pageref{lastpage}} \pubyear{}

\maketitle

\label{firstpage}

\begin{abstract}
We use the direct Fourier method to calculate the redshift-space power spectrum of the maxBCG cluster catalog \citep{2007astro.ph..1265K} -- currently by far the largest existing galaxy cluster sample. The total number of clusters used in our analysis is $12,616$. After accounting for the radial smearing effect caused by photometric redshift errors and also introducing a simple treatment for the nonlinear effects, we show that currently favored low matter density ``concordance'' $\Lambda$CDM cosmology provides a very good fit to the estimated power. Thanks to the large volume ($\sim 0.4 \,h^{-3}\,\rm{Gpc}^3$), high clustering amplitude (linear effective bias parameter $b_{\rm{eff}}\sim 3\times(0.85/\sigma_8)$), and sufficiently high sampling density ($\sim 3 \times 10^{-5} \,h^{3}\,\rm{Mpc}^{-3}$) the recovered power spectrum has high enough signal to noise to allow us to find weak evidence ($\sim 2\sigma$ CL) for the acoustic features. These results are encouraging in light of the several proposed large cluster surveys. In case we use the photometric redshift errors as suggested in \citet{2007astro.ph..1265K} we are left with the excess large-scale power which has previously been noticed by several other authors.  
\end{abstract}


\section{Introduction}
Since the flight of the {\sc Cobe} satellite in the early 1990's the field of observational cosmology has witnessed extremely rapid progress which has culminated in the establishment of the Standard Model for cosmology -- the ``concordance'' model \citep{1999Sci...284.1481B,2003ApJS..148..175S}. This progress has been largely driven by the precise measurements of the angular temperature fluctuations of the Cosmic Microwave Background (CMB) \citep{2000ApJ...545L...5H,2002ApJ...571..604N,2003ApJS..148....1B,2003MNRAS.341L..23G,2003ApJ...591..556P}. However, in order to break several degeneracies inherent in the CMB measurements one has to complement this data with other sources of information, such as the measurements of the SNe~Ia luminosity distances \citep{1998AJ....116.1009R,1999ApJ...517..565P} or with the measurements of the large-scale structure (LSS) of the Universe as traced by galaxies or galaxy clusters. 

The simplest descriptor one can extract from the LSS measurements is the matter power spectrum. The broad-band shape of this spectrum is sensitive to the shape parameter $\Gamma = \Omega_mh$, and thus is useful in helping to establish the currently favored low matter density ``concordance'' model, as well as the amount of baryons in relation to the total matter $f_b=\Omega_b/\Omega_m$. Currently the two largest galaxy redshift surveys are the 2dF Galaxy Redshift Survey\footnote{http://www.mso.anu.edu.au/2dFGRS/} (2dFGRS) and the Sloan Digital Sky Survey\footnote{http://www.sdss.org/} (SDSS) with its latest data release 5 (DR5), providing redshifts to $\sim 220,000$ and $\sim 675,000$ galaxies, respectively. The SDSS galaxy sample consists of two broad classes: (i) the MAIN galaxy sample, reaching redshifts of $z \sim 0.25$; (ii) the Luminous Red Galaxy (LRG) sample covering redshift range $z \sim 0.15 - 0.5$. Some other characteristics of these surveys are listed in Table \ref{tab}.  

\begin{table*}
\caption{Some characteristics of the 2dFGRS, SDSS DR5, and maxBCG samples.}
\label{tab}
\begin{tabular}{l|l|l|l|l|l}
Survey & Number of & Sky area & Redshift & Comoving volume & Effective bias parameter wrt to\\
 & objects & (deg$^2$) & coverage & ($h^{-3}\,\rm{Gpc}^3$) & the model with $\sigma_8=0.85$\\
\hline
\hline
2dFGRS & $\sim 220,000$ & $\sim 1200$ & $\lesssim 0.3$ & $\sim 0.07$ & $\sim 1.0$ \\
\hline
SDSS DR5 & $\sim 675,000$ & $\sim 5700$ & MAIN: $\lesssim 0.25$ & MAIN: $\sim 0.2$ & MAIN: $\sim 1.2$\\
& & & LRG: $\sim 0.15 - 0.5$ & LRG: $\sim 1.3$ & LRG: $\sim 2.0$\\
\hline
maxBCG & $\sim 13,800$ & $\sim 7000$ & $ 0.1 - 0.3$ & $\sim 0.4$ & $\sim 3$
\end{tabular}
\end{table*}

In addition to the precise measurement of the broad-band shape of the power spectrum, SDSS LRG and 2dFGRS samples have proven useful in allowing the detection of theoretically predicted oscillatory features in the spectrum. The source of these spectral fluctuations is the same as that giving rise to the prominent peak structure in the CMB angular power spectrum -- acoustic oscillations in the tightly coupled baryon-photon fluid prior to the epoch of recombination \citep{1970Ap&SS...7....3S,1970ApJ...162..815P}. However, the relative level of fluctuations in the matter power spectrum is strongly reduced compared to the corresponding fluctuations in the CMB angular spectrum, owing to the fact that most of the matter in the Universe is provided by the cold dark matter (CDM) that does not participate in acoustic oscillations. After recombination, or more precisely after the end of the so-called ``drag-epoch'' which happens at redshifts of a few hundred, as the baryons are released from the supporting pressure of the photon gas they start to fall back to the CDM potential wells that have started to grow already since the matter-radiation equality. Since the amount of baryons in the total matter budget is not completely negligible -- baryons make up roughly one fifth -- the acoustic structure imprinted onto the baryonic component also gets transferred to the spatial distribution of the CDM. After baryonic and CDM components have relaxed one ends up with $\sim 5\%$ relative fluctuations in the total mater power spectrum.

In order to observe these relatively small fluctuations one needs to have:
\begin{enumerate}
\item Large redshift surveys, to reduce cosmic variance;
\item High enough spatial sampling density, to decrease discreteness noise.
\end{enumerate}
Both of these criteria are currently best met by the SDSS LRG sample, and indeed, at the beginning of 2005 the detection of the acoustic peak in the spatial two-point correlation function was announced by \citet{2005ApJ...633..560E}. At the same time the final power spectrum measurements of the 2dFGRS also revealed the existence of the acoustic features \citep{2005MNRAS.362..505C}. Baryonic acoustic oscillations (BAO) in the power spectrum of the SDSS LRG (DR4) sample were found by \citet{2006A&A...449..891H} and the corresponding cosmological parameter estimation was carried out in \citet{2006A&A...459..375H}. Also, several other detections of BAO have been reported: \citet{2006astro.ph..5302P} and \citet{2007MNRAS.374.1527B} used SDSS photometric LRG sample, \citet{2006PhRvD..74l3507T} analyzed spectroscopic LRG sample, and finally \citet{2007ApJ...657..645P} carried out a combined analysis of the SDSS MAIN and LRG spectroscopic galaxy samples.

The usefulness of BAO arises from the fact that it can provide us with a standard ruler in the form of the size of the sound horizon at decoupling\footnote{To be more precise again, at the end of the ``drag-epoch''. In the ``concordance'' $\Lambda$CDM model the sound horizon at the ``drag-epoch'' is $\sim 5\%$ larger compared to the one at recombination.}, enabling us to carry out a purely geometric cosmological test: comparing the apparent size of the ruler along and perpendicular to the line of sight with the physical size of the sound horizon (which can be well calibrated from the CMB data) one is able to find Hubble parameter $H(z)$ and angular diameter distance $d_A(z)$ corresponding to the redshift $z$ (e.g. \citealt{2003PhRvD..68f3004H}). Obviously, having determined $H(z)$ and $d_A(z)$ one can put constraints on the dark energy (DE) equation of state parameter $w_{DE}(z)$. In that respect radial modes, which enable us to find $H(z)$, are more useful since $H(z)$ is related to $w_{DE}(z)$ through a single integration whereas $d_A(z)$, which is given by the angular modes, involves double integration over $w_{DE}(z)$. Thus one would certainly benefit from accurate redshift information.
  
If one wishes to use the BAO signal as a precise standard ruler there are a few complications one has to overcome:
\begin{enumerate}
\item As the density field goes nonlinear couplings between Fourier modes will modify the BAO signal, leading to the damping of the oscillations;
\item In order to fully exploit the information in galaxy/cluster surveys one has to understand the relation of these objects to the underlying matter density field, i.e. one has to have a realistic model for biasing. 
\end{enumerate}
These two complications can be fully addressed only through costly N-body simulations. There are several papers that have investigated the detectability and possible systematics of BAO extraction using the numerical simulations e.g. \citet{1999MNRAS.304..851M,2005Natur.435..629S,2005ApJ...633..575S,2007APh....26..351H,2007A&A...462....7K,2007astro.ph..2543A,2007astro.ph..3620S}. However, significantly more work towards these directions is probably needed before we can harness the full power provided by the potentially very clean geometric tool in the form of BAO in the matter power spectrum.

It is worth pointing out that actually one of the first possible hints for the existence of BAO in the matter power spectrum came from the analysis of the spatial clustering of the Abell/ACO clusters \citep{2001ApJ...555...68M}.\footnote{Interesting features in the spatial two-point correlation function of the Abell/ACO cluster sample were also revealed by \citet{1997MNRAS.289..801E}.} The power spectrum measurement in \citet{2001ApJ...555...68M} used $637$ clusters with richness $R \geq 1$. In this paper we analyze the maxBCG cluster sample \citep{2007astro.ph..1265K} which is currently by far the largest cluster catalog available, spanning a redshift range $z=0.1-0.3$, containing $13,823$ objects, and covering $\sim 0.4 \,h^{-3}\,\rm{Gpc}^3$ of comoving volume (see Table \ref{tab} for comparison with 2dFGRS and SDSS DR5). The biggest advantage of using galaxy clusters instead of the typical, i.e. $L_*$-galaxies, is the fact that their spatial clustering signal is strongly amplified with respect to the clustering of the underlying matter, and thus one can achieve the same signal to noise measurement of the power spectrum by using correspondingly less number of objects.     
\begin{figure}
\centering
\includegraphics[width=\plotwd]{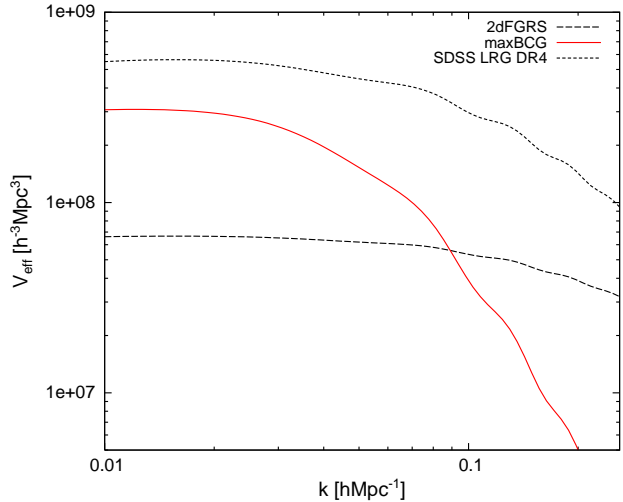}
\caption{The effective volume (see Eq. (\ref{eq2})) of the maxBCG cluster sample in comparison to the 2dFGRS and SDSS LRG DR4 samples.}
\label{fig1}
\end{figure}
In Fig. \ref{fig1} we have compared the effective volume of the maxBCG survey with the two other surveys that have lead to the detection of BAO: 2dFGRS and SDSS LRG. Under the assumption of Gaussianity the power spectrum measurement errors $\Delta P$ are simply found as: $\Delta P/P = 2/V_{\rm{eff}}/V_k$, where $V_k$ is the volume of the shell in $k$-space over which the angular average is taken, i.e. $V_k = 4 \pi k^2 \Delta k/(2\pi)^3$. It is no surprise that the SDSS LRG sample is currently unbeatable. However, we notice that maxBCG sample should provide better measurement of the power spectrum on scales larger than $k \sim 0.1 \,h\,\rm{Mpc}^{-1}$ compared to the 2dFGRS. The rapid fall of the effective volume of the maxBCG sample on smaller scales is caused by the photometric redshift errors of $\delta z \sim 0.01$ as estimated by \citet{2007astro.ph..1265K}.

Thus one would expect the maxBCG sample to reveal acoustic features. As our further analysis shows this indeed turns out to be the case, albeit with a relatively modest confidence level.

Our paper is organized as follows. In Section 2 we give a brief description of the maxBCG catalog with the corresponding selection effects. The core part of this paper, which is devoted to the power spectrum analysis, is given in Section 3. Finally, Section 4 contains our conclusions.

\section{Data and survey selection function}\label{sec2}
A maxBCG cluster catalog\footnote{http://umsdss.physics.lsa.umich.edu/catalogs/\\maxbcg\textunderscore public\textunderscore catalog.dat} \citep{2007astro.ph..1265K} is compiled via the ``red sequence'' cluster detection method \citep{1998AJ....116.2644O,2000AJ....120.2148G,2003ApJS..148..243B} applied to the SDSS photometric data. The catalog contains $13,823$ galaxy clusters with velocity dispersions greater than $\sim 400$ km/s and covers $\sim 7000$ square degrees of sky between redshifts $0.1$ and $0.3$. The photometric redshifts are estimated using the tight relation between the ridgeline color and redshift and are claimed to have accuracy $\delta z \equiv \sqrt{\left < (z_{\rm photo} -z_{\rm spec})^2\right >} \sim 0.01$ essentially independent of redshift. For more details about the catalog we refer to the original source \citet{2007astro.ph..1265K}.

\begin{figure}
\centering
\includegraphics[width=\plotwd]{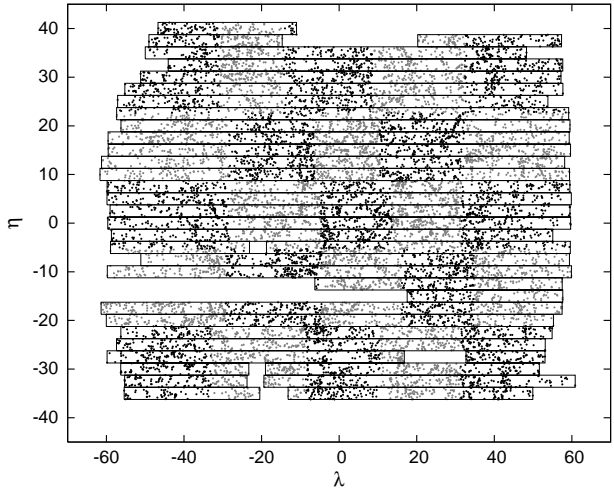}
\caption{Angular distribution of $12,616$ maxBCG clusters in SDSS survey coordinates $(\eta,\lambda)$. The $5 \times 5$ chessboard pattern of black/gray points shows the angular part of the division used for ``jackknife'' error analysis. The union of solid rectangles represents our reconstruction of the angular mask.}
\label{fig2}
\end{figure}

In our power spectrum analysis we are going to neglect the three southern SDSS stripes, leaving us with $12,616$ galaxy clusters over $\sim 6800$ square degrees of sky. With the redshift range $0.1 < z < 0.3$ this corresponds to $\sim 0.4 \,h^{-3}\,\rm{Gpc}^3$ of comoving cosmic volume. Neglecting the three narrow southern stripes helps us to achieve a better behaved survey window function with reduced sidelobes, and thus with reduced leakage of power. 

The angular distribution of the remaining clusters is given in Fig. \ref{fig2}. Here the angular coordinates are plotted using the SDSS survey coordinates $(\eta,\lambda)$ (e.g. \citealt{2002AJ....123..485S}). The chessboard pattern of black/gray points in Fig. \ref{fig2} represents the division of the sky into $5 \times 5$ angular regions used for ``jackknife'' error analysis. We also use three divisions along the redshift direction, making the total of $3\times 5 \times 5 =75$ regions, each containing approximately $168$ clusters. The union of solid rectangles in Fig. \ref{fig2} represents our reconstruction of the survey angular mask. All the rectangles are aligned with the SDSS imaging scan stripes. As the number density of galaxy clusters in the maxBCG sample is rather high, one can determine relatively accurately the beginning, ending, and also possible gaps in the scan stripes. We call the angular mask obtained this way a ``minimal'' mask in contrast to the ``maximal'' mask, which is built in the same manner except that each of the rectangles is extended by an amount corresponding to the mean cluster separation. We carry out our power spectrum analysis using both of these angular masks.   

\begin{figure}
\centering
\includegraphics[width=\plotwd]{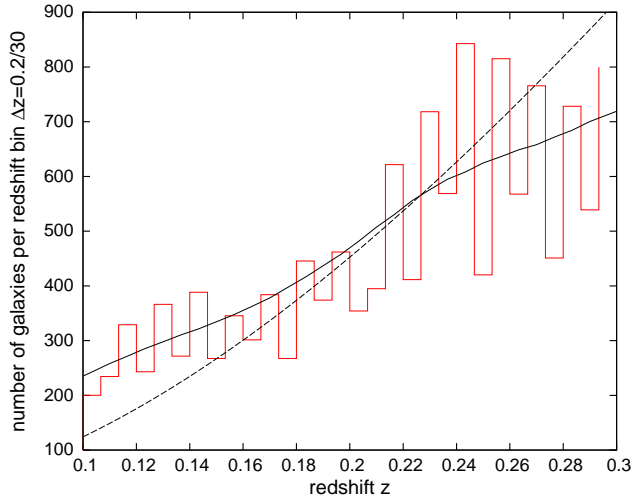}
\caption{Redshift histogram of the maxBCG sample. The solid smooth line corresponds to the cubic spline fit and the dashed line represents the expected number of clusters per redshift bin for the volume-limited sample.}
\label{fig3}
\end{figure}

In Fig. \ref{fig3} we show the redshift distribution of maxBCG clusters. Here the solid smooth line corresponds to the cubic spline fit and the dashed line represents the expected number of clusters per redshift bin $\Delta z = 0.2/30$ for the volume-limited sample\footnote{In \citet{2007astro.ph..1265K} it is suggested that maxBCG sample is rather close to being volume-limited.} assuming a spatially flat $\Lambda$CDM cosmology with $\Omega_m=0.27$. To convert this redshift distribution to the radial selection function $\bar{n}(r)$, i.e. the comoving number density of clusters at distance $r$, we fix the background cosmology the low matter density model mentioned above. Our power spectrum calculations are done for both of the radial selection models shown in Fig. \ref{fig3}.   

For the full survey selection function $\bar{n}(\mathbf{r})$ we assume, as usual, that it can be factorized to the product of angular and radial parts, i.e. $\bar{n}(\mathbf{r})=\bar{n}(\mathbf{\hat{r}})\bar{n}(r)$ ($r\equiv |\mathbf{r}|$, $\mathbf{\hat{r}}\equiv \mathbf{r}/r$). Here $\bar{n}(r)$ is the radial selection function and the angular selection part $\bar{n}(\mathbf{\hat{r}})$ is assumed to take value $1$ if the unit direction vector $\mathbf{\hat{r}}$ is inside the survey mask and $0$ otherwise, i.e. we assume a simple binary angular mask.

\section{Power spectrum analysis}
We calculate the redshift-space power spectrum of the maxBCG catalog using the direct Fourier method as described in \citet{1994ApJ...426...23F} (FKP). Strictly speaking, power spectra determined in this way are so-called pseudospectra, meaning that the estimates derived are convolved with a survey window. Since in the case of the analyzed maxBCG sample the volume covered is very large, reaching $0.4 \,h^{-3}\,\rm{Gpc}^3$, and also the survey volume has relatively large dimensions along all perpendicular directions, the correlations in the Fourier space caused by the survey window are rather compact. On intermediate scales and in the case the power spectrum binning is chosen wide enough, FKP estimator gives a good approximation to the true underlying power.

Instead of direct summation as presented in \citet{1994ApJ...426...23F} we speed up the calculations using Fast Fourier Transforms (FFTs). This gives rise to some extra complications. As our density field, which is built using the Triangular Shaped Cloud (TSC) \citep{1988csup.book.....H} mass assignment scheme, is now ``restricted to live'' on a regular grid with a finite cell size, we have to correct for the smoothing effect this has caused. Also, if our underlying density field contains spatial modes with higher frequency than our grid's Nyquist frequency, $k_{\rm{Ny}}$, then these will be ``folded back'' into the frequency interval the grid can support, increasing power close to $k_{\rm{Ny}}$ -- the so-called aliasing effect. Quite often in the literature this aliasing effect is not properly accounted for, which is only fine in cases where the grid's Nyquist frequency used in the analysis is significantly higher than the spatial frequencies where one wants to measure the spectrum. We correct for the aliasing effect using the iterative scheme as described in \citet{2005ApJ...620..559J} with a slight modification: we do not approximate the small-scale spectrum by a simple power law, but also allow for the possible running of the spectral index, i.e. the parametric shape of the power spectrum is taken to be a parabola in log--log. 

For the full details of our power spectrum calculation method along with several tests we refer the reader to \citet{2006A&A...446...43H,2006A&A...449..891H}. As a very brief summary, our spectrum determination consists of the following steps:
\begin{itemize}
\item Determination of the survey selection function, i.e. mean expected number density without any clustering $\bar{n}(\mathbf{r})$ (see Section \ref{sec2}). This smooth field, with respect to which the fluctuations are measured, is modeled using a random catalog that contains $100$ times more objects than the real maxBCG catalog, i.e. $1,261,600$ in total;
\item Calculation of the overdensity field on a $512^3$-grid using TSC mass assignment scheme;
\item FFT of the gridded overdensity field;
\item Calculation of the raw 3D power spectrum by taking the modulus squared from the output of the previous step;
\item Subtraction of the shot noise component from the raw 3D spectrum;
\item Recovery of the angle averaged spectrum using a modified version of the iterative scheme of \citet{2005ApJ...620..559J}.    
\end{itemize}

\begin{figure}
\centering
\includegraphics[width=\plotwd]{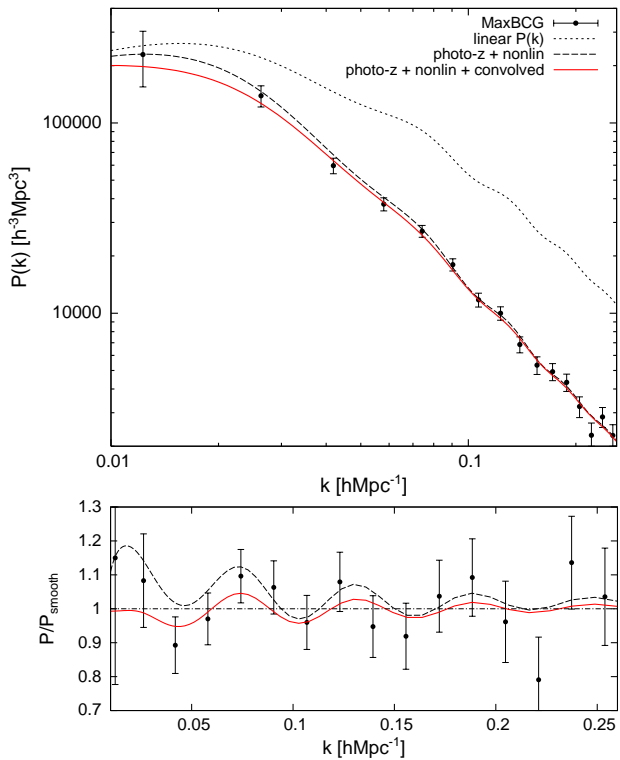}
\caption{Upper panel: Filled circles with error-bars show the redshift-space power spectrum of maxBCG cluster sample. Solid line represents the best fitting model in case of three free parameters: $b_{\rm{eff}}$, $q$, $\sigma$, and assumes ``jackknife'' error model. Long dashed line shows the same model without survey window convolution applied and short dashed line corresponds to the linear model spectrum. Lower panel: the same as above, with the broad-band shape of the spectrum removed by dividing with a ``smooth'' model spectrum without BAO. Here we do not show linear model as it falls way out of this small plotted region.}
\label{fig4}
\end{figure}

\begin{figure}
\centering
\includegraphics[width=\plotwd]{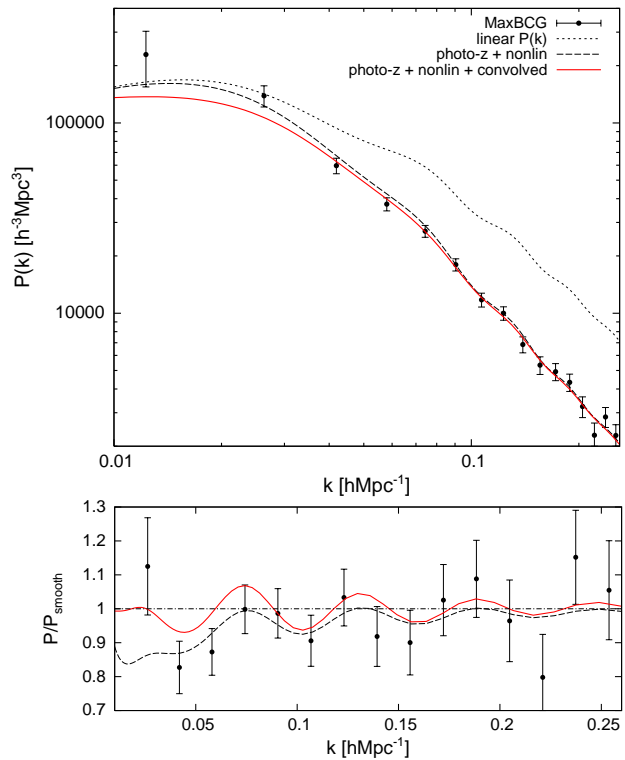}
\caption{Same as Fig. \ref{fig4}, with the only difference that here we have two free parameters: $b_{\rm{eff}}$, $q$, and the value for $\sigma$ is fixed to $30\,h^{-1}\,\rm{Mpc}$.}
\label{fig5}
\end{figure}

The results of our power spectrum calculation along with fitted models are shown in Figs. \ref{fig4} and \ref{fig5}. Here the error-bars $\Delta P$ are calculated as described in FKP, which assumes that the density field follows Gaussian statistics (see also \citealt{1998ApJ...499..555T}):
\begin{equation}
\frac{\Delta P}{P} = \sqrt{\frac{2}{V_{\rm{eff}}V_k}}\,,
\end{equation}
where the effective volume $V_{\rm{eff}}$ (see Fig. \ref{fig1}) is given by:
\begin{equation}\label{eq2}
V_{\rm{eff}} = \Omega\cdot\int\limits_{r_{\min}}^{r_{\max}}\left[ \frac{\bar{n}(r)P(k)}{1+\bar{n}(r)P(k)}\right]^2r^2\,{\rm d}r\,, 
\end{equation}
and $V_k$ is the volume of the $k$-space shell over which the angular average of the 3D spectrum is taken, i.e.
\begin{equation}
V_k=\frac{4\pi k^2 \Delta k}{(2\pi)^3}\,.
\end{equation}
In Eq. (\ref{eq2}) $r_{\min}$ and $r_{\max}$ are comoving distances corresponding to the redshift bounds of the survey ($0.1$ and $0.3$ in case of maxBCG), $\bar{n}(r)$ is the radial selection function, and $\Omega$ is the area of the sky in steradians covered by the survey.

\begin{figure}
\centering
\includegraphics[width=\plotwd]{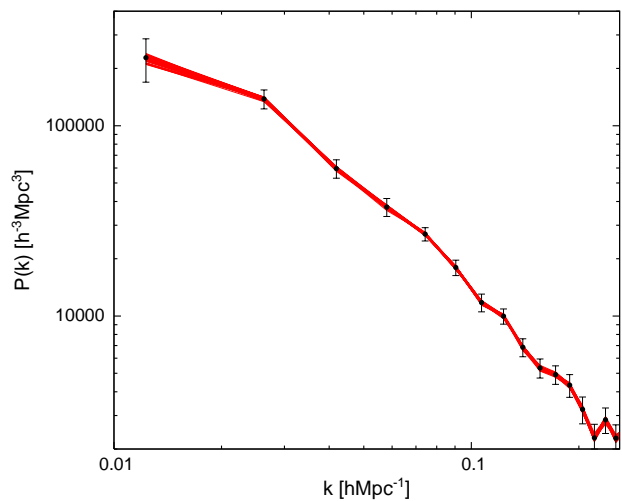}
\caption{Piecewise linear lines show 75 ``jackknife'' spectra and filled circles with error-bars represent the corresponding errors derived from the variability of these spectra.}
\label{fig6}
\end{figure}

In addition to the FKP errors we have also estimated the power spectrum variance via the ``jackknife'' method (see e.g. \citealt{1993stp..book.....L}). For this purpose we have divided the survey volume into $5 \times 5$ angular regions (see Fig. \ref{fig2}) along with $3$ redshift bins, making the total of $75$ regions, each containing on average $168$ clusters. Now the power spectrum is calculated $75$ times with each time one of the regions omitted. The covariance of the power spectrum can now be estimated via the scatter of the power spectrum measurements: 
\begin{eqnarray}
\rm{cov}(P_i,P_j) & = & \frac{N-1}{N}\cdot\sum\limits_{n=1}^{N}\left( P_i^{(n)}-\bar{P_i}\right)\left( P_j^{(n)}-\bar{P_j}\right)\,,\\
\bar{P_i} & = & \frac{1}{N}\cdot \sum\limits_{n=1}^{N}P_i^{(n)}\,.
\end{eqnarray}
Here $P_i^{(n)}$ represents the power spectrum measurement for the $i$-th power spectrum bin in case the $n$-th survey subvolume was excluded from the calculations. All the $75$ spectra along with the inferred ``jackknife'' error-bars for each of the spectrum bins $i$, $\Delta P_i=\sqrt{\rm{cov}(P_i,P_i)}$, are shown in Fig. \ref{fig6}. 
\begin{figure}
\centering
\includegraphics[width=\plotwd]{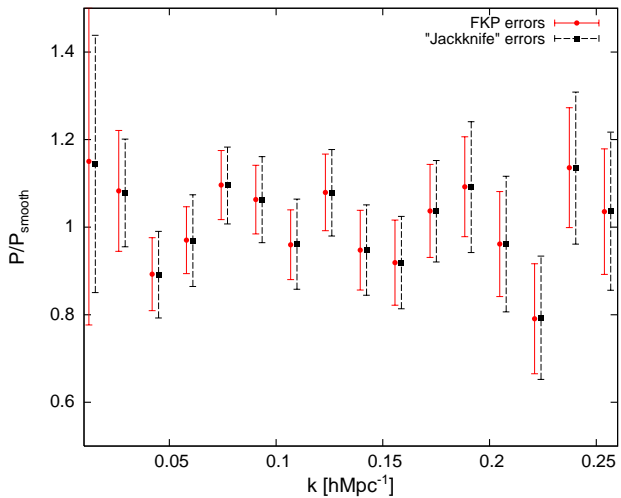}
\caption{Comparison of the FKP and ``jackknife'' error estimates.}
\label{fig7}
\end{figure}
``Jackknife'' and FKP error-bars are compared in Fig. \ref{fig7} where for clarity we have removed the smooth broad-band component of the spectrum. We see that both errors are in good agreement. On the largest scales the ``jackknife'' errors probably slightly underestimate the real errors, whereas on smaller scales they might be more appropriate as they do not rely on the Gaussianity assumption. However, this small difference does not influence our subsequent analysis, i.e. both error estimates lead to fully consistent results.   

Before we can fit model spectra to the data there are several effects one has to take into account:
\begin{itemize}
\item Photometric redshift errors lead to significant damping of the spectrum;
\item Our observed spectrum is a pseudospectrum and thus before fitting one has to convolve model spectra with the survey window;
\item On smaller scales nonlinear effects become noticeable and one needs a model to account for these;
\item And finally, even without photometric redshift errors there are spectral modifications caused by the redshift-space distortions.
\end{itemize}
Our model for the relation between the linear spectrum $P_{\rm{lin}}(k)$ and the observed one $P_{\rm{obs}}(k)$ that tries to accommodate all the four effects listed above reads as:
\begin{equation}\label{eq6}
P_{\rm{obs}}(k) = \int {\rm d}k'\,k'^2P(k')K(k',k)\,,
\end{equation}
where
\begin{equation}\label{eq7}
P(k) = b_{\rm{eff}}^2\left(1+qk^{\frac{3}{2}}\right)f(k)P_{\rm{lin}}(k)\,.
\end{equation}
Here $b_{\rm{eff}}$ is the cluster bias parameter that also incorporates the boost of the power at large scales due to the linear redshift-space distortions. Smaller scale redshift distortions are fully eliminated if the cluster detection algorithm is able to perform a perfect finger-of-God compression. However, in reality the compression is not ideal and thus there should be some extra uncertainties for the central redshift of the galaxy cluster. This can be modelled as an additional smoothing effect in addition to the photometric redshift errors of the individual galaxies. The factor $(1+qk^{3/2})$ is our simplistic model to treat effects due to nonlinear evolution. This form is very similar to the one suggested in \citet{2005MNRAS.362..505C}, however in \citet{2006A&A...459..375H} we found that the power law index $3/2$ provides better fit to the Halo Model\footnote{For a comprehensive review on Halo Model see \citet{2002PhR...372....1C}.} results compared to the value $2$ used in \citet{2005MNRAS.362..505C}. The function $f(k)$ in Eq. (\ref{eq7}) models the damping of the spectrum due to photometric redshift (photo-z) errors. Under the flat sky approximation it is easy to derive the analytic form for $f(k)$ which reads:
\begin{equation}
f(k)=\frac{\sqrt{\pi}}{2\sigma k}\rm{erf}(\sigma k)\,,
\end{equation}
where the error function $\rm{erf}(x)\equiv\frac{2}{\sqrt{\pi}}\int_0^{x}\exp(-t^2)\,{\rm d}t\,$. This result assumes that photo-z errors follow Gaussian distribution with dispersion $\delta z$ and $\sigma$ is the corresponding spatial smoothing scale, i.e. $\sigma = \frac{c}{H_0}\delta z\,$. And finally, Eq. (\ref{eq6}) models the convolving effect of the survey window. We find that the mode coupling kernels $K(k',k)$ can be well described by the following analytic fit:
\begin{equation}
K(k',k)=K(k,k')=\frac{C}{kk'}\left[g(k+k')-g(k-k')\right]\,,
\end{equation}
where
\begin{equation}
g(x)=\arctan\left(\frac{b^4+2a^2x^2}{b^2\sqrt{4a^4-b^4}}\right)\,,
\end{equation}
and the normalization constant $C$ is derived by demanding that $\int K(k,k')k'^2\,\rm{d}k'\equiv 1$, giving:
\begin{equation}
C=\frac{1}{\pi b \sqrt{2-\left(\frac{b}{a}\right)^2}}\,.
\end{equation}
In the case of the maxBCG survey geometry the best fitting parameters $a$ and $b$ are found to be: $a=0.0044$ and $b=0.0045$. For the motivation of this analytic form see \citet{2006A&A...449..891H}.

\begin{figure}
\centering
\includegraphics[width=\plotwd]{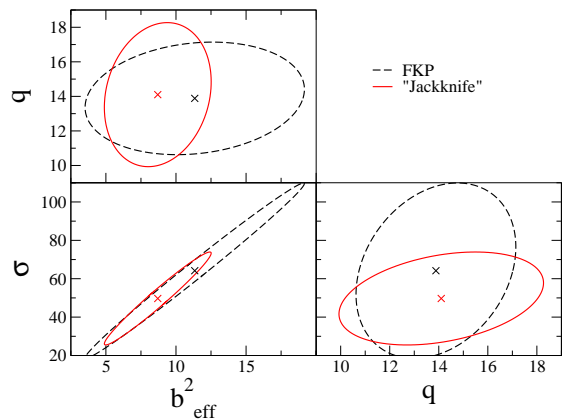}
\caption{Error ellipses for the case with three free parameters. The solid and dashed lines correspond to the ``jackknife'' and FKP error models, respectively. The crosses mark the best fitting parameter values. Note the strong degeneracy between $b_{\rm{eff}}^2$ and $\sigma$.}
\label{fig}
\end{figure}

In this paper we keep the background cosmology fixed to the ``concordance'' flat $\Lambda$CDM model with $\Omega_m=0.27$, $\Omega_b=0.045$, and $h=0.7$. The bias parameters quoted below are measured with respect to the model with $\sigma_8=0.85$. Our decision not to fit for the cosmological parameters is based on two main reasons:
\begin{enumerate}
\item Photo-z uncertainties that are given in \citet{2007astro.ph..1265K} are tested only for the case of the brightest cluster members that have spectroscopic redshifts available. The errors for the whole galaxy population in clusters might differ, and be possibly larger than the numbers given in \citet{2007astro.ph..1265K};
\item Uncertainties in the cluster finding algorithm which are hard to quantify and which might lead to the additional smoothing/overmerging, and thus to the extra uncertainties in derived cluster redshifts.
\end{enumerate}
Any untreated systematics in the determination of the radial smoothing scale $\sigma$ will start to interfere with the cosmological parameter estimation, resulting in biased estimates. Due to the uncertainties listed above we first treat the radial smoothing scale $\sigma$ as a free parameter. Thus we are left with three free parameters: $b_{\rm{eff}}$, $\sigma$, $q$. We also investigate separately the case where the radial smoothing scale is fixed at $c/H_0 \cdot \delta z \simeq 30\,h^{-1}\,\rm{Mpc}$ as suggested in \citet{2007astro.ph..1265K}. Some of the results of these calculations are presented in Figs. \ref{fig4} and \ref{fig5}, and in Tables \ref{tab1} and \ref{tab2}.  
\begin{table*}
\caption{Best fitting model parameters and $\chi^2$ values along with the inferred confidence levels for BAO detection for the case with three free parameters: $b_{\rm{eff}}$, $q$, $\sigma$.}
\label{tab1}
\begin{tabular}{|l|c|c|c|c|c|c|}
\hline
 $P(k)$ & \multicolumn{6}{c|}{3 free parameters: $b_{\mathrm{eff}}^2$, $q$, $\sigma$; $16-3=13$ dof; expected $\chi^2\simeq 13.0 \pm 5.1$}\\
 \cline{2-7}
 error model & $b_{\mathrm{eff}}^2$ & $q$ & $\sigma$ & $\chi_{\mathrm{wiggly}}^2$ & $\chi_{\mathrm{smooth}}^2$ & BAO detection \\
\hline
FKP & $11.3 \pm 5.2$ & $13.9 \pm 2.1$ & $64 \pm 30$ & $7.48$ & $10.9$ & $1.8\sigma$ \\
``Jackknife'' & $8.7 \pm 2.5$ & $14.1 \pm 2.7$ & $50 \pm 16$ & $9.51$ & $14.2$ & $2.2\sigma$\\
\hline
\end{tabular}
\end{table*}
\begin{table*}
\caption{Analog of Table \ref{tab1} for the case with two free parameters: $b_{\rm{eff}}$, $q$, and the value for $\sigma$ is fixed to $30\,h^{-1}\,\rm{Mpc}$.}
\label{tab2}
\begin{tabular}{|l|c|c|c|c|c|}
\hline
 $P(k)$ & \multicolumn{5}{c|}{$\sigma=30$; 2 free parameters: $b_{\mathrm{eff}}^2$, $q$; $16-2=14$ dof; expected $\chi^2\simeq 14.0 \pm 5.3$}\\
 \cline{2-6}
 error model & $b_{\mathrm{eff}}^2$ & $q$ & $\chi_{\mathrm{wiggly}}^2$ & $\chi_{\mathrm{smooth}}^2$ & BAO detection \\
\hline
FKP & $5.70 \pm 0.32$ & $11.7 \pm 1.9$ & $11.8$ & $14.5$ & $1.6\sigma$ \\
``Jackknife'' & $5.76 \pm 0.37$ & $11.9 \pm 2.3$ & $12.5$ & $15.6$ & $1.8\sigma$\\
\hline
\end{tabular}
\end{table*}
To calculate theoretical model spectra we have used fitting formulae for the transfer functions presented in \citet{1998ApJ...496..605E}. In the lower panels of Figs. \ref{fig4} and \ref{fig5} we have divided the spectra by the ``smooth'' model spectrum with acoustic oscillations removed (see \citealt{1998ApJ...496..605E} for details). For fitting purposes we have used Levenberg-Marquardt  $\chi^2$-minimization technique as described in \citet{1992nrfa.book.....P} with the additional modification to allow for the nondiagonal data covariance matrices. We perform the fitting for both error models: (i) FKP errors with diagonal covariance matrix, (ii) full ``jackknife'' covariance matrix. The solid lines in Figs. \ref{fig4} and \ref{fig5} represent the best fitting ``wiggly'' models for the case of ``jackknife'' errors. In these figures we also demonstrate the damping effect of the window convolution. The corresponding linear spectra are shown with short dashed lines. One can see that the shape of the measured spectrum deviates strongly from the linear model spectrum, which is mainly driven by the photo-z errors. The best fitting parameter values along with the corresponding $\chi^2$ values are given in Tables \ref{tab1}, \ref{tab2}. There we also present the $\chi^2$ value for the best fitting ``smooth'' model. In general we can see that the results for the $\chi^2$-statistic agree well with expectations $13.0 \pm 5.1$ and $14.0 \pm 5.3$ for $13$ and $14$ degrees of freedom, relevant for Table \ref{tab1} and \ref{tab2}, respectively. Comparing the $\chi^2$ values for the best fitting ``smooth'' and ``wiggly'' models one can assess the confidence level for the BAO detection, the results of which are presented in the last columns of Table \ref{tab1} and \ref{tab2}. We note that using the full ``jackknife'' covariance matrix we obtain higher confidence for the detection of the oscillatory features in the spectrum. This is easy to understand: our first error model, where we assume diagonal covariance matrix and FKP errors is certainly not realistic. Survey window, and more importantly nonlinear mode-mode coupling induces significant correlations between neighboring power spectrum bins. Stronger correlations amongst the bins make the spectrum ``harder to distort'', and thus any surviving oscillatory features gain higher statistical weight. It is interesting to note that in case we allow $\sigma$ to be a free parameter the best fitting values are certainly larger than $30\,h^{-1}\,\rm{Mpc}$. Also the bias parameter $b_{\rm{eff}}$ has a noticeable increase compared to the $\sigma = 30\,h^{-1}\,\rm{Mpc}$ case. In fact there is a strong degeneracy between $b_{\rm{eff}}$ and $\sigma$: if one chooses to increase $b_{\rm{eff}}$, one can still obtain a good fit to the data by also increasing $\sigma$ by a suitable amount (see Fig. \ref{fig}). The fact that the case with three free parameters favors higher value for $\sigma$ (and thus also for $b_{\rm{eff}}$) is mainly driven by the relatively large amplitudes of the two first power spectrum bins. On the other hand, if one assumes that $\sigma = 30\,h^{-1}\,\rm{Mpc}$ is a reasonably correct value then one is left with some discrepancy on the largest scales between the model and data (see Fig. \ref{fig5}). It is interesting that some large-scale ``extra power'' has also been detected by several other authors e.g. \citet{2006astro.ph..5302P,2007MNRAS.374.1527B} who analyzed SDSS photometric LRG sample.\footnote{See also Fig. 4 in \citealt{2006A&A...449..891H} where SDSS spectroscopic LRG sample reveals some mild excess of power on the largest scales.} Of course these mild discrepancies on the largest scales might be due to inaccurate treatment of the survey selection effects. Taking into account several uncertainties concerning the appropriate value for $\sigma$, and also uncertainties concerning the survey selection effects, we are here unfortunately unable to draw any firm conclusions about the existence of the excess large-scale power. This issue certainly deserves a dedicated study and currently the best available sample for this purpose is probably the SDSS LRG spectroscopic sample.

However, one extra exercise we can perform here is to try to investigate how well the obtained values for the bias parameter $b_{\rm{eff}}$ agree with the model expectations. If one assumes a flat $\Lambda$CDM model with the parameters as given above (i.e. $\Omega_m=0.27$, $\sigma_8=0.85$) then for the mass-limited cluster survey covering the same redshift range and sky area as maxBCG sample, and giving the same number of clusters, one infers a lower mass limit of $M_{\rm{low}}=7.1\times10^{13}\,h^{-1}M_{\odot}$. We have calculated $M_{\rm{low}}$ using the Sheth-Tormen mass function \citep{1999MNRAS.308..119S} averaged over the past light-cone (see \citealt{2006A&A...446...43H} for details). The corresponding light-cone averaged effective bias parameter including increase of power due to large-scale redshift-space distortions (see again \citealt{2006A&A...446...43H}) with respect to the $z=0$ linear spectrum turns out to be $b_{\rm{eff}}\simeq 2.8$ ($b_{\rm{eff}}^2 \simeq 7.8$).\footnote{In reality of course for the mass-limited sample the number density drops with redshift, while maxBCG catalog is close to being volume-limited. We have calculated that to achieve a comoving number density of $3.2 \times 10^{-5} \,h^{3}\,\rm{Mpc}^{-3}$, as relevant for the maxBCG catalog, the lower mass limits for redshifts $0.1$ and $0.3$ should be $7.8\times10^{13}\,h^{-1}M_{\odot}$ and $6.6\times10^{13}\,h^{-1}M_{\odot}$, respectively. The effective bias parameters with respect to the $z=0$ linear matter power spectrum for these redshifts and lower mass limits are $b_{\rm{eff}}^{z=0.1}\simeq 2.75$ and $b_{\rm{eff}}^{z=0.3}\simeq 2.78$, i.e. basically independent of redshift, and also agreeing very well with the light-cone averaged value of $2.8$ quoted in the main text. This approximate constancy of the clustering amplitude at different redshifts for the objects with fixed comoving number density is a well known result.} Comparing this value of $b_{\rm{eff}}^2\simeq 8$ with the ones given in Tables \ref{tab1} and \ref{tab2} one sees that $\sigma = 30\,h^{-1}\,\rm{Mpc}$ case has somewhat lower bias value, whereas the case with free $\sigma$ tends to have values for $b_{\rm{eff}}^2$, which seem to agree reasonably well (especially true for the case with ``jackknife'' errors). Thus this simple analysis would favor the ``extra smoothing'' hypothesis over the ``extra power'' one.

\begin{figure}
\centering
\includegraphics[width=\plotwd]{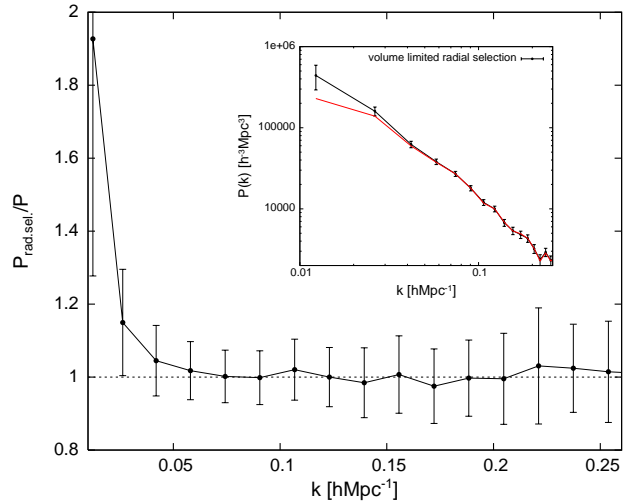}
\caption{Stability test of the power spectrum measurement with respect to changes in the radial selection  function. The solid line with error-bars in the inset of the figure shows the measured power spectrum in case the radial selection corresponds to the volume-limited survey. The other solid line displays our ``default'' spectrum measurement. To increase the visibility of differences we have divided by the ``default'' spectrum in the main panel of the figure.}
\label{fig8}
\end{figure}
\begin{figure}
\centering
\includegraphics[width=\plotwd]{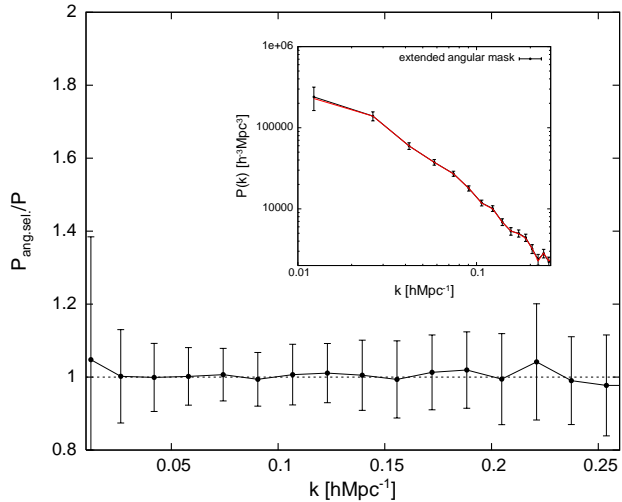}
\caption{Analog of Fig. \ref{fig8} for the stability test with respect to the variations in the angular mask.}
\label{fig9}
\end{figure}

As a final part of our power spectrum analysis we test the stability of our results against possible uncertainties in the survey selection function. For this purpose we have repeated power spectrum calculations with modified radial and angular selections. In the inset of Fig. \ref{fig8} we show the change induced in the power spectrum measurement if one replaces the smooth cubic spline fitted radial selection with the one corresponding to the volume-limited sample (see also Fig. \ref{fig3}). For clarity the main panel of this figure shows the spectrum divided with our ``default'' spectrum. We see that at almost all the scales, except for the very largest ones, both spectra agree very well with each other. However, this exercise shows that the largest scales are very vulnerable with respect to the unaccounted slow trends in the survey selection. The results of the similar exercise for the angular selection function are shown in Fig. \ref{fig9}. Here the agreement between the power spectrum measurements using the ``minimal'' and ``maximal'' angular masks (see Section \ref{sec2} for explanation) is almost ideal. Thus we can conclude that our results seem to be robust with respect to the uncertainties in the survey selection, which is especially true for scales with wavenumbers $k\gtrsim 0.03 \,h\,\rm{Mpc}^{-1}$.  

\section{Conclusions and discussion}
We have calculated the redshift-space power spectrum of the maxBCG cluster sample \citep{2007astro.ph..1265K} -- currently the largest cluster catalog in existence. After correcting for the radial smoothing caused by the photometric redshift errors, treating the convolution with a survey window, and introducing a simple model for the nonlinear effects we show that ``concordance'' $\Lambda$CDM model is capable of providing very good fit to the data. Moreover, we are able to find weak evidence for the acoustic features in the spatial clustering pattern of the maxBCG sample. If radial smoothing scale $\sigma$, effective bias parameter $b_{\rm{eff}}$, and nonlinear distortion parameter $q$ are treated as completely free parameters the model with acoustic oscillations is favored by $2.2\sigma$ over the corresponding ``smooth'' model without any oscillatory behavior. The evidence for BAO weakens somewhat ($1.8\sigma$), and we are also left with some ``extra power'' on large scales, if we fix the smoothing scale $\sigma$ to $30\,h^{-1}\,\rm{Mpc}$ as suggested by the photometric redshift errors $\delta z = 0.01$ estimated in \citet{2007astro.ph..1265K}. We note that these photometric errors actually apply only for the very brightest cluster members and in reality the scatter for the whole sample might be larger. Also the cluster finding algorithm itself might introduce additional smoothing/overmerging along the line of sight. Due to these uncertainties we have decided not to carry out any cosmological parameter study in this paper, as any untreated systematics will be immediately propagated to the parameter estimates, leading to biased results. We have estimated power spectrum errors using two different methods: (i) FKP errors, (ii) ``jackknife'' errors, obtained by dividing the survey into $5\times 5 \times 3$ chunks containing equal number of galaxies. Both of these errors turn out to agree very well with each other. We have also shown the stability of our power spectrum measurement with respect to the possible uncertainties in the radial and angular selection function.

The detectability of BAO with $\sim 10^4$ galaxy clusters is very encouraging result keeping in mind the future cluster surveys, such as the ones based on the measurement of the thermal Sunyaev-Zeldovich effect \citep{1972CoASP...4..173S}, e.g. {\sc Spt}\footnote{http://spt.uchicago.edu/}, {\sc Planck}\footnote{www.rssd.esa.int/Planck/}, or the proposed $100,000$-cluster X-ray survey eROSITA\footnote{www.mpe.mpg.de/erosita/MDD-6.pdf}. In reality the main target of these cluster surveys is to map the evolution of the cluster number density as a function of redshift, since this quantity is very sensitive to the amplitude and to the growth rate of the density perturbations, and as such it provides a powerful tool to constrain the properties of DE. The possibility to also find the clustering power spectrum can be seen as a useful byproduct of these surveys that comes essentially ``for free''. However, both cluster number count study and spatial clustering analysis needs as an input estimates for the cluster redshifts, with the former probably doing relatively well with rather poor redshift estimates (e.g. photometric redshifts). This is a rather complicated task since one arguably needs of order $10$ galaxy redshifts per cluster to reliably infer cluster redshift. In that respect for the measurement of BAO it is certainly less costly to sample the cosmic density field using e.g. LRGs or blue emission line galaxies. The latter type of objects are targets for the currently ongoing WiggleZ project \citep{2007astro.ph..1876G} at AAO that attempts to measure redshifts for $400,000$ objects over $\sim 1000$ deg$^2$ and cover the redshift range $0.5< z < 1$. There is also a {\sc Wfmos} instrument (see e.g. \citealt{2005A&G....46e..26B}) construction planned for the Gemini and Subaru observatories that will hopefully be completed in 2012. This new multi-object spectrograph will be able to measure the spectra of $\sim 5000$ objects at the same time. There are plans to perform a wide field ($\sim 2000$ deg$^2$) redshift survey giving spectra for $\sim 2 \times 10^6$ galaxies up to redshifts of $z \sim 1.3$ together with a narrower ($\sim 200$ deg$^2$) and deeper ($z \sim 2 - 3$) survey with a yield of $\sim 5 \times 10^5$ galaxies. For the more distant future (around 2020) one would expect a superb measurement of the matter power spectrum using $\sim 10^9$ galaxy redshifts obtainable via the 21 cm measurements by the Square Kilometre Array ({\sc Ska})\footnote{http://www.skatelescope.org/} (e.g. \citealt{2004NewAR..48.1063B,2005MNRAS.360...27A}). However, before {\sc Wfmos} and {\sc Ska} become available several wide area imaging surveys, such as Pan-STARRS\footnote{pan-starrs.ifa.hawaii.edu/} or Dark Energy Survey\footnote{https://www.darkenergysurvey.org/} ({\sc Des}), will be performed. Although these photometric surveys lack accurate redshift information, the huge number of objects detectable over large sky areas significantly compensate this shortcoming, allowing us to obtain a highly competitive measurement of BAO \citep{2005MNRAS.363.1329B}.

\section*{Acknowledgments}
I thank Ofer Lahav for useful comments and suggestions and Jonathan Braithwaite for carefully reading the manuscript.
I acknowledge the support provided through a PPARC/STFC postdoctoral fellowship at UCL and travel support from ETF grant 7146.

Funding for the SDSS and SDSS-II has been provided by the Alfred P. Sloan Foundation, the Participating Institutions, the National Science Foundation, the U.S. Department of Energy, the National Aeronautics and Space Administration, the Japanese Monbukagakusho, the Max Planck Society, and the Higher Education Funding Council for England. The SDSS Web Site is http://www.sdss.org/.

The SDSS is managed by the Astrophysical Research Consortium for the Participating Institutions. The Participating Institutions are the American Museum of Natural History, Astrophysical Institute Potsdam, University of Basel, University of Cambridge, Case Western Reserve University, University of Chicago, Drexel University, Fermilab, the Institute for Advanced Study, the Japan Participation Group, Johns Hopkins University, the Joint Institute for Nuclear Astrophysics, the Kavli Institute for Particle Astrophysics and Cosmology, the Korean Scientist Group, the Chinese Academy of Sciences (LAMOST), Los Alamos National Laboratory, the Max-Planck-Institute for Astronomy (MPIA), the Max-Planck-Institute for Astrophysics (MPA), New Mexico State University, Ohio State University, University of Pittsburgh, University of Portsmouth, Princeton University, the United States Naval Observatory, and the University of Washington.

\bibliographystyle{mn2e}
\bibliography{aamnem99,references}

\bsp

\label{lastpage}

\end{document}